\documentclass
[preprint,pra,amsfonts,amssymb,10pt,showpacs,showkeys,byrevtex,onecolumn]{revtex4}%
\usepackage{amsmath}
\usepackage{amsfonts}
\usepackage{amssymb}
\usepackage{graphicx}%
\providecommand{\U}[1]{\protect\rule{.1in}{.1in}}

\begin{document}
\title{Relaxation, frequency shifts and other phenomena at the \\transition between diffusion and ballistic motion.}
\author{Christopher Swank, Robert Golub}
\affiliation{}
\affiliation{}

\begin{abstract}
\bigskip There are many fields where the transition from diffusive to
ballistic motion is important. Here we deal with relaxation processes in nmr
in gases. Correlation functions for trajectory variables (position and
velocity) valid across this transition are known for several geometries in the
case of specular wall reflections. In this work we show that the conditional
probability density $p\left(  \overrightarrow{r},t|\overrightarrow{r}%
_{0},t_{0}\right)  $ for a random walk satisfies the telegrapher's equation
and present an analytic solution for this function. We will use it for
calculating the relaxation due to an axion mediated force and a magnetic
dipole impurity.

\end{abstract}
\maketitle
\tableofcontents

Authors' address:

Physics Department

North Carolina State University

Raleigh, NC 27695-8202

\section{Introduction}

The influence of magnetic field inhomogeneities on nmr measurements has been
discussed by many authors. It is usually sufficient to treat the problem using
second order perturbation theory, (Redfield theory, \cite{Redf}, \cite{Slich})
according to which frequency shifts and relaxation rates can be shown to
depend on the frequency spectrum of the magnetic field fluctuations seen by
the ensemble of particles moving through the inhomogeneous field. This
spectrum is given by the Fourier transform of the correlation functions of the
various field components.

Usually the field variations are assumed to be described by constant gradients
so the correlation functions of the field components are proportional to the
correlation functions of the particle positions. McGregor \cite{Mcgreg} has
shown how to calculate the correlation functions using diffusion theory.
Recently in \cite{Barab} it has been shown that these trajectory correlation
functions can also be generalized for conditions (long mean free path for gas
collisions) where diffusion theory is not valid and a general form has been
given valid for all values of mean free path.

The correlation functions are calculated by averaging the two successive
positions, $\overrightarrow{x}\left(  t\right)  ,\overrightarrow{x^{\prime}%
}(t^{\prime})$ over a joint probability distribution which in turn is
calculated in terms of the conditional probability distribution, $p\left(
\overrightarrow{x},t|\overrightarrow{x^{\prime}},t^{\prime}\right)  $ which
can be calculated by diffusion theory when that is valid \cite{Mcgreg}. By
making use of the form of the conditional probability distribution calculated
by diffusion theory it has been possible to apply these ideas to fields with
arbitrary spatial variation \cite{Clayt} and the technique has been applied to
placing a limit on a possible axion mediated unconventional force which falls
off exponentially with distance from the walls of the container \cite{Petu}.

In this work we extend the work on the generalized correlation functions to
calculate a generalized joint probability distribution valid for all values of
mean free path. The generalized conditional probability is shown to satisfy a
partial differential equation. This allows the study of the transition between
diffusion and ballistic motion for the case of arbitrary field variation and
should be applicable to a wide range of physical problems.

In \cite{Barab} the authors show that the velocity autocorrelation of confined
gasses including gas collisions (wall collisions are assumed specular) can be
expressed as a weighted sum of solutions to the stochastic damped harmonic
oscillator and appropriate boundary conditions. They solved for the velocity
correlation function of gasses confined in cylindrical vessels , and used its
spectrum to predict a frequency shift important in searches for particle
electric dipole moments \cite{JMP}, \cite{LamGo} . \ In \cite{Swank} the
authors applied the same idea to the rectangular cell combined with the
observation that specular reflections enable a correlation in 2 dimensions to
be decoupled into two correlations of 1 dimension. In \cite{GolRelax} the
authors derive the position autocorrelation function starting with the
previously determined velocity autocorrelation.

\section{From ballistic to diffusive motion \ }

\subsection{An equation for the conditional probability density}

\qquad\ \ A free (collision free) particle starting at $x_{0}$ with velocity
$v$ has a conditional probability density given by
\begin{equation}
p(\overrightarrow{x},t~|~\overrightarrow{x_{0}},t_{0}=0)=\delta
(\overrightarrow{x_{o}}+\overrightarrow{v}t-\overrightarrow{x}) \label{1}%
\end{equation}

Equation (\ref{1}) satisfies the wave equation as it is a function of $\left(
x-vt\right)  .$ In three dimensions%
\begin{equation}
\frac{\partial^{2}p(\overrightarrow{x},t~|~\overrightarrow{x_{0}},t_{0}%
)}{\partial t^{2}}-v^{2}\triangledown^{2}p(\overrightarrow{x}%
,t~|~\overrightarrow{x_{0}},t_{0})=0
\end{equation}

\bigskip To include the effects of collisions we introduce a damping term as
in \cite{Barab} and write%
\begin{equation}
\frac{\partial^{2}p(\overrightarrow{x},t~|~\overrightarrow{x_{0}},t_{0}%
)}{\partial t^{2}}+\frac{1}{\tau_{c}}\frac{\partial p(\overrightarrow
{x},t~|~\overrightarrow{x_{0}},t_{0})}{\partial t}-v^{2}\triangledown
^{2}p(\overrightarrow{x},t~|~\overrightarrow{x_{0}},t_{0})=0 \label{3}%
\end{equation}

This is known as the Telegrapher's equation. It has applications to the
propagation of sound in an absorbing medium, to a damped vibrating string,
among others, and provides a method for eliminating the infinite propagation
velocity from the theory of heat conduction or diffusion.

For large $\tau_{c}$ (\ref{3}) goes over to the wave equation, for small
$\tau_{c}$ it becomes the diffusion equation. This allows us to study the
transition region for intermediate values of $\tau_{c}.$ Similarly for short
times when the variation is rapid the second time derivative dominates showing
that the short time motions is ballistic going over to diffusive motion for
longer times when $\tau_{c}$ is short enough.

There is an extensive literature on the applications of this equation. We can
only give some entry points to this. In reference \cite{Haji}, the authors
have taken a similar approach to ours in order to calculate the Green's
function in a bounded region,\ The equation has been applied, among others, to
the conduction of heat, since it overcomes the problem of the infinite
propagation velocity of the Fourier theory \cite{MandF} and has been called
the relativistic theory of heat conduction \cite{ALi1}. It is claimed to be
applicable to heat conduction for distances on the order of or less than the
collision mean free path \cite{Haji2}, but this is contradicted by
\cite{Chen}. Chen's argument is specific to heat conduction and does not apply
to the random walk problem considered here. \ A review of the applications of
different forms of the equation to heat conduction is given in \cite{Joseph}.
Goldstein \cite{Gold}, as early as 1951, in a little noticed paper, has
discussed the Telegrapher's equation in relation to the random walk problem.

\ \ \ It has also been applied to the transmission of light through opaque
materials \cite{Lem}, \cite{Dur}.

Here we apply it to the generic random walk problem and show how it can be
used to study relaxation and frequency shifts in nmr. We show by comparing to
simulations that our results apply for short distances and times

\subsection{A generalized conditional probability density}

Equation (\ref{3}) can be solved by the usual separation of variables. Writing
the space-time dependent function as
\begin{equation}
F(\overrightarrow{k},t)e^{i\overrightarrow{k}\cdot\overrightarrow{x}}%
\end{equation}

we have%
\begin{equation}
\frac{\partial^{2}F(\overrightarrow{k},t)}{\partial t^{2}}+\frac{1}{\tau_{c}%
}\frac{\partial F(\overrightarrow{k},t)}{\partial t}+k^{2}v^{2}%
F(\overrightarrow{k},t)=0 \label{4}%
\end{equation}
The solution is a linear combination of terms with time dependence given by
\begin{equation}
\exp\left(  -\eta_{1,2}t\right)
\end{equation}
where%
\begin{align}
\eta_{1,2}  &  =\frac{1}{2\tau_{c}}\left(  1\pm s\right) \\
s_{{}}  &  =\sqrt{1-4\omega^{2}\tau_{c}^{2}}\\
\omega &  =kv
\end{align}

\bigskip

We require a solution of (\ref{4}) that satisfies the boundary conditions%
\begin{equation}
\overrightarrow{n}\cdot\overrightarrow{\triangledown}p(\overrightarrow
{x},t~|~\overrightarrow{x_{0}},t_{0})_{\delta\Omega}=0,
\end{equation}

yields the known diffusion theory result for $p(\overrightarrow{x}%
,t~|~\overrightarrow{x_{0}},t_{0})$ in the limit \ of small $\tau_{c},$

and yields a correlation function in agreement with the results of ref.
\cite{GolRelax}, equation (26), where we now specialize to one dimension for
simplicity.%
\begin{align}
R_{xx}\left(  \tau\right)   &  =\left\langle \overrightarrow{x(\tau)}%
\cdot\overrightarrow{x(0)}\right\rangle ,\\
&  =\int\overrightarrow{x(\tau)}\cdot\overrightarrow{x_{0}}p(\overrightarrow
{x},t~|~\overrightarrow{x_{0}},t_{0})p\left(  \overrightarrow{x_{0}}%
,t_{0}\right) \label{0}\\
&  =\frac{8}{\pi^{2}}v^{2}\tau_{c}\sum_{n=1,3,5..}\frac{1}{n^{2}s_{n}}\left[
\frac{e^{-\eta_{2}\tau}}{\eta_{2}}-\frac{e^{-\eta_{1}\tau}}{\eta_{1}}\right]
\\
&  ==\frac{4L^{2}}{\pi^{4}}e^{-\frac{t}{2\tau_{c}}}\sum\frac{1}{n^{4}}\left\{
\cosh\left(  \frac{s_{n}}{2\tau_{c}}t\right)  +\frac{1}{s_{n}}\sinh\left(
\frac{s_{n}}{2\tau_{c}}t\right)  \right\}
\end{align}
\ \ 

where%

\begin{align}
s_{n}  &  =\sqrt{1-4\omega_{n}^{2}\tau_{c}^{2}}\\
\eta_{1,2}  &  =\frac{1}{2\tau_{c}}\left(  1\pm s_{n}\right) \label{00}\\
\omega_{n}  &  =\frac{n\pi v}{L}%
\end{align}
and $p\left(  \overrightarrow{x_{0}},t_{0}\right)  $ is the position
probability distribution at time $t=t_{0}.$

Notice, for the ballistic limit, $\underset{\tau_{c}\rightarrow\infty}{\lim
}\left(  s_{n}\right)  =i\infty.~$and $\underset{\tau_{c}\rightarrow\infty
}{\lim}\left(  \frac{s_{n}}{2\tau_{c}}\right)  =i\omega_{n}$ while for small
$\tau_{c},$ $s_{n}\rightarrow1-2\omega_{n}^{2}\tau_{c}^{2}$ and
\begin{align}
\frac{s_{n}}{2\tau_{c}}  &  \rightarrow\frac{1}{2\tau_{c}}-\omega_{n}^{2}%
\tau_{c}\\
\omega_{n}^{2}\tau_{c}  &  =\left(  \frac{n\pi v}{L}\right)  ^{2}\tau_{c}%
\sim\left(  \frac{n\pi}{L}\right)  ^{2}D\sim\frac{1}{\tau_{D}}%
\end{align}
where $\tau_{D}$ is the time to diffuse a distance $L$.

\bigskip In addition $p(x,t~|~x_{0},0)$ must satisfy the initial conditions%
\begin{align}
p(x,t  &  =0|~x_{0},0)=\delta\left(  x-x_{0}\right) \\
\left.  \frac{\partial p(x,t|~x_{0},0)}{\partial t}\right\vert _{t=0}  &  =0
\end{align}

Strictly speaking the conditional probability distribution used here and in
\cite{Mcgreg} which satisfies the homogeneous equation with an initial
condition equal to $\delta\left(  \overrightarrow{r}\right)  $ is not exactly
the same as the Green's function which is the solution for a point impulse
source. In the case of a linear equation containing only a first order time
derivative it can be shown that the Green's function does have the required
initial condition so that in the case of diffusion the functions are
identical. This does not hold in the present case.

The solution to (\ref{3}) satisfying these criterion (in one dimension,
$-L/2\leq x\leq L/2$) is%
\begin{align}
p(x,t~|~x_{0},0)  &  =\frac{1}{L}\left\{  \sum_{n=even}\cos\left(  \frac{n\pi
x}{L}\right)  \cos\left(  \frac{n\pi x_{0}}{L}\right)  +\sum_{n=odd}%
\sin\left(  \frac{n\pi x}{L}\right)  \sin\left(  \frac{n\pi x_{0}}{L}\right)
\right\}  \times..\nonumber\\
&  \times\left\{  \cosh\left(  \frac{s_{n}t}{2\tau_{c}}\right)  +\frac
{1}{s_{n}}\sinh\left(  \frac{s_{n}t}{2\tau_{c}}\right)  \right\}  e^{-\frac
{t}{2\tau_{c}}} \label{5}%
\end{align}
We can see that this is a solution by considering the trigonometric functions
in exponential form. Then every term will be of the form%
\begin{equation}
\exp\left[  \pm i\frac{n\pi x}{L}-\eta_{1,2}t\right]
\end{equation}
so the equation is satisfied. \ The boundary and initial conditions
are\ satisfied. The other requirements are shown to be satisfied by direct
calculation. We take the sums over $\pm$ integers so as to include both
directions of motion.

The initial condition $p(x,t_{0}|x_{0},t_{0})=$ $\delta(x-x_{0})~\ $is seen to
hold since the inclusion of the terms multiplied by $\frac{1}{s_{n}}~$\ does
not effect the initial condition, they are identically zero at $t=0.$

This solution applies to particles moving with a single velocity. It is
necessary to average over the appropriate velocity distribution. The
periodicity of the solutions means that reflections from the walls are
properly taken into consideration (see appendix).

This solution is seen to be correct in the limits of large and small $\tau
_{c}.$ The generalization to 2 and 3 dimensions is obvious.

Correlation functions of arbitrary functions of position are seen to reduce
immediately \ to combinations of the Fourier transforms of those arbitrary functions.

In order to check its validity in the intermediate region we have performed
Monte-Carlo simulations. The results are compared with equation (\ref{5}) and
the agreement is seen to be excellent.

\section{Comparison with Monte\ Carlo simulations}

We introduce dimensionless variables%
\begin{align}
x^{\prime}  &  =\frac{x}{L}\\
t^{\prime}  &  =\frac{t}{\tau_{c}}%
\end{align}

Then equation (\ref{3})\bigskip\ can be written%
\begin{equation}
\frac{\partial^{2}p(\overrightarrow{x^{\prime}},t^{\prime}~|~\overrightarrow
{x_{0}^{\prime}},t_{0}^{\prime})}{\partial t^{\prime2}}+\frac{\partial
p(\overrightarrow{x^{\prime}},t^{\prime}~|~\overrightarrow{x_{0}^{\prime}%
},t_{0}^{\prime})}{\partial t^{\prime}}-l_{c}^{2}\triangledown^{2}%
p(\overrightarrow{x},t~|~\overrightarrow{x_{0}},t_{0})=0
\end{equation}
where $l_{c}=v\tau_{c}/L=\lambda_{c}/L$ is the normalized collision mean free
path. In these units the velocity $v^{\prime}=l_{c}$. We consider the
solutions in the region $-0.5<x^{\prime}<0.5$ and drop the primes from now on.

The conditional density is simulated in one dimension. This is accomplished by
tracking an ensemble of trajectories , $N=500,000$, for a given initial
position $x_{0}$ and velocity $\pm v.$ The trajectories scatter at random
according to the cumulative distribution function of a collision in time,%

\begin{align*}
P_{scatter}  &  =1-e^{-t/\tau_{c}}\\
t_{scatter}  &  =-\tau_{c}\ln(1-P_{scatter})
\end{align*}
where the probability of scattering, $P_{scatter},$ is a uniform distribution
between $0,1$. At the start of each trajectory, or directly after a scattering
event, a time of next scattering is selected by randomly selecting
$P_{scatter}$. $\;$At the time of a scattering event there is a 50/50 chance
of it being scattered in the reverse direction with speed $v,~$versus
maintaining its current trajectory. This is consistent with isotropic
scattering in one dimension. \ Wall reflections are treated as specular.
\ \ The whole treatment is easily modified to allow for the transport mean
free path with non-isotropic scattering in higher dimensions.

Below is a plot (fig.1) for $l^{\prime}=0.05.$ This is in the region where
diffusion theory is valid, however unlike the present theory the diffusion
theory does not account for the peaks representing the unscattered particles
moving with finite velocity.%

\begin{figure}
[ptb]
\begin{center}
\includegraphics[
height=3.0386in,
width=4.8199in
]%
{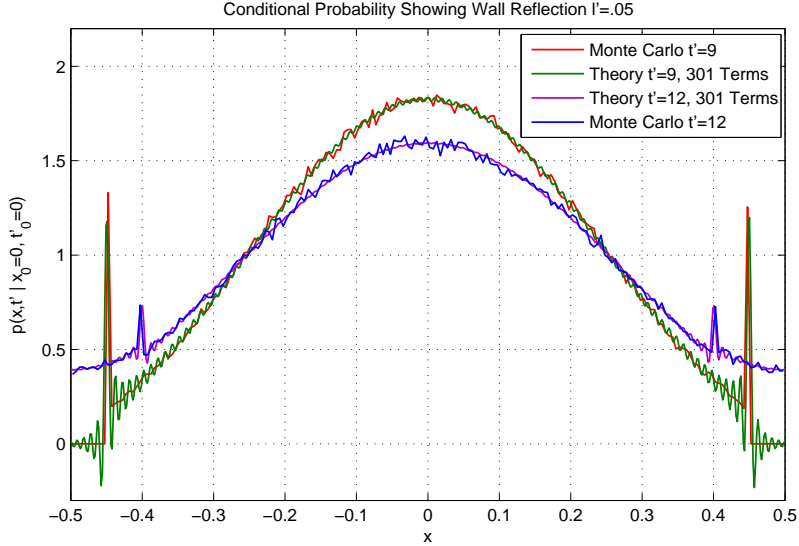}%
\caption{Conditional probability density with $x_{0}=0$ ,for the case
$l_{c}=0.05,$ well into the diffusion limit, showing the agreement between the
analytic result and the simulations for $t^{\prime}=9$ and $12$. The peaks
representing the unscattered particles are clearly visible at these times as
is the Gibbs phenomenon due to cutting off the high frequency terms.}%
\end{center}
\end{figure}

The Gibbs phenomena is clearly seen in this plot. The Gibbs phenomenon arises
due to the impossible task of representing an abrupt change with a finite
series of continuous functions. In the finite sum the abrupt change can only
be approximated by the fastest oscillating component, and in the limit of an
infinite series the fastest frequency is infinite and the exact form is
recovered. \ However for purposes of plotting we cannot sum to infinite terms,
and must accept the rapid oscillations, or smooth the rapid oscillations for a
more pleasant and informative picture. To smooth the conditional density we
take a running 4 point average. This is done for $4\times500$ points across
the region. Typically $20,001$ terms are summed prior to smoothing the
functions but in this figure (fig. 2) we show the sum over only 301 terms for visibility.%

\begin{figure}
[ptb]
\begin{center}
\includegraphics[
height=3.0395in,
width=4.4168in
]%
{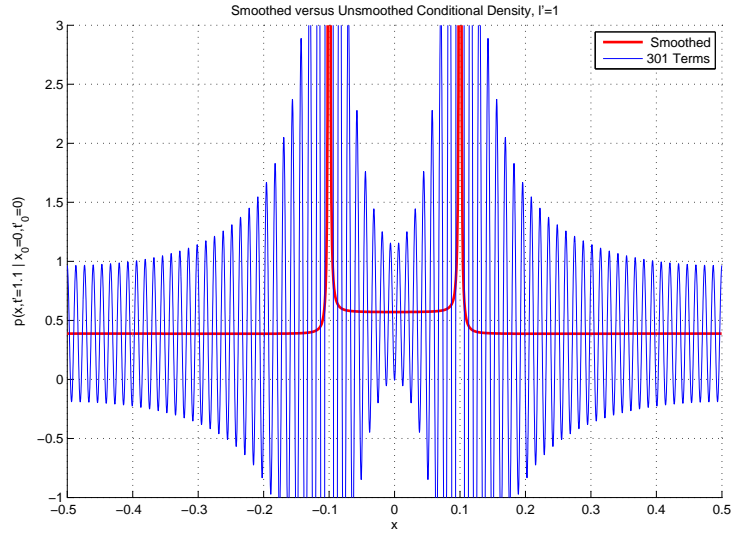}%
\caption{Conditional probability density with $x_{0}=0$ ,for the case
$l_{c}=1,t^{\prime}=1.1$, approaching the ballistic limit. The particles have
made one wall collision and have passed through each other $\left(  v^{\prime
}=l_{c}\right)  .$ Shown are a truncated sum $(N=301$ terms$)$ and the result
of smoothing a sum with N=20,001.}%
\end{center}
\end{figure}

\ \qquad The smoothing correctly shows the average of the rapid oscillations
and gives a clearer picture. \ The following (fig. 3) is a plot of the
conditional density with $l^{\prime}=0.3.$ This is in the transition region.
The peaks representing the initial trajectory of the particles are seen to be
large for longer than a collision time. It is seen that the theory agrees
extremely well with the Monte-Carlo simulations.%

\begin{figure}
[ptb]
\begin{center}
\includegraphics[
height=3.0395in,
width=4.5948in
]%
{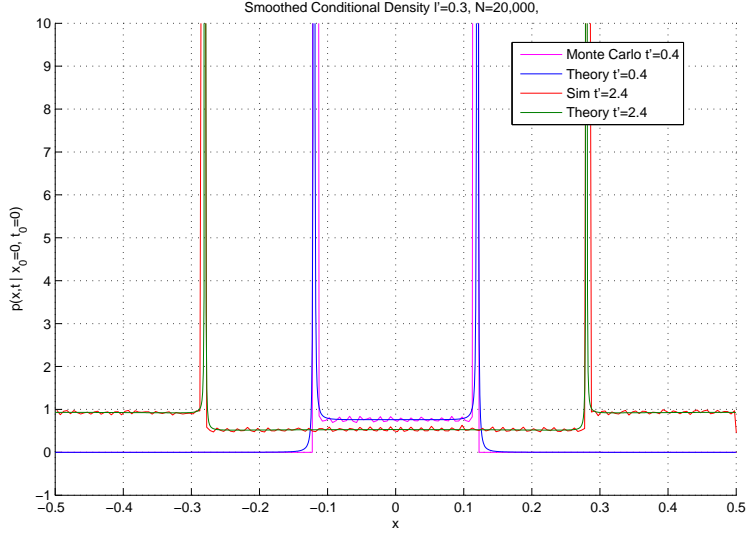}%
\caption{Conditional probability density with $x_{0}=0$ ,for the case
$l_{c}=.3,t^{\prime}=0.4$ and $2.4.$ A wall collision has taken place between
the two times shown. The build up of the wake of scattered particles is
clearly visible.}%
\end{center}
\end{figure}

In the earlier time the trajectories are moving away from each other, in the
later time they have reflected against the wall and are moving toward each
other. The initial trajectories are scattering at a slow enough rate to leave
a uniform wake of probability behind them, the uniform probability is
deposited in layers behind the traversing peaks. That is to say the rate of
change of probability going from the initial trajectory into the wake is slow
compared to the rate of crossing the cell. The wake appears uniform and the
curve reminiscent of diffusion is not seen. \ The following \ (fig.4)is a plot
for $l^{\prime}=1\bigskip.$\ More toward the ballistic region, however still
somewhat in the transition region.\bigskip%
\begin{figure}
[ptb]
\begin{center}
\includegraphics[
height=3.0386in,
width=4.1087in
]%
{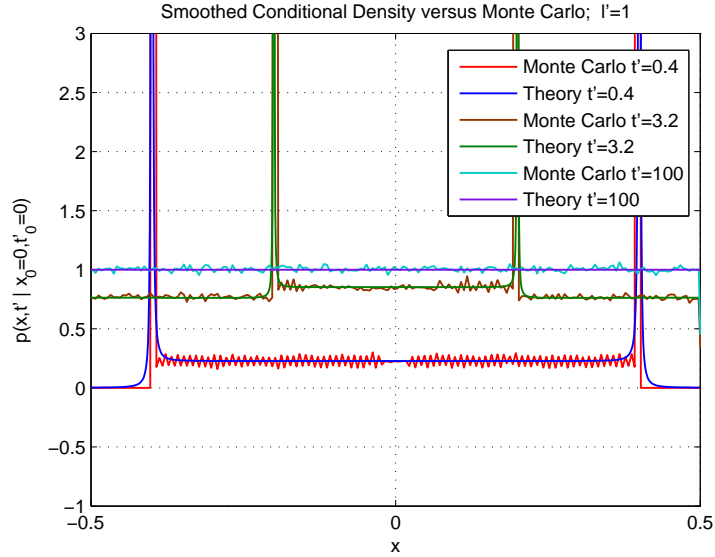}%
\caption{Conditional probability density with $x_{0}=0$ ,for the case
$l_{c}=1,t^{\prime}=0.4,3.2$ and $100.$ By the later time the density is
uniform and the unscattered peaks are gone,}%
\end{center}
\end{figure}

This behavior is similar to the $l^{\prime}=0.3$ case. The wake of probability
deposited behind the peaks is less, however after a long time the peaks can no
longer be seen and a uniform distribution is found. The proposed theory is
shown to agree very well with the Monte-Carlo simulations.

We see in general that as we are considering particles starting with
velocities $\pm v$, there are 2 peaks, representing unscattered particles
moving away from $x_{0}$ in opposite directions. The peaks are decaying in
time while a 'wake' of scattered particles builds up in the region behind
them. The time development can be followed as the particles make multiple
collisions with the walls. For short $\tau_{c}$ the wake morphs into the
diffusion theory result, where the 'wake' is position dependent, after some
time. For the ballistic case of longer $\tau_{c}$, the wake does not develop
any significant spatial dependence while the peaks decay with time. In all
cases the wake behind the unscattered particles are larger than the wake ahead
of them.

\section{Calculation of Spectrum of Correlation functions.}

Now, starting with (\ref{5}) we can calculate the position correlation
function according to (\ref{0}) and then take the Fourier transform to obtain
the spectrum. We can differentiate (\ref{5}) with respect to $\tau$ to get the
velocity position correlation function according to, \cite{Pap}
\begin{equation}
R_{xv}\left(  \tau\right)  =-\frac{\partial}{\partial\tau}R_{xx}\left(
\tau\right)  \label{01}%
\end{equation}
and a further differentiation will give the velocity autocorrelation function,
\cite{Barab}. We can calculate correlation functions involving quantities
which vary arbitrarily with position. The results will be valid for short
times and small distances allowing to consider fields varying over distances
small compared to the collisional mean free path.

As it is usually the spectrum of the correlation function that is of interest
we give here the Fourier transform of (\ref{5}) that can then be integrated
over arbitrary field distributions to give the desired spectrum.%
\begin{align}
G(x,x_{0},\omega) &  =\int_{0}^{\infty}p(x,\tau~|~x_{0},0)e^{i\omega\tau}%
d\tau\\
&  =\frac{1}{L}\left\{  \sum_{n=even}\cos\left(  \frac{n\pi x}{L}\right)
\cos\left(  \frac{n\pi x_{0}}{L}\right)  +\sum_{n=odd}\sin\left(  \frac{n\pi
x}{L}\right)  \sin\left(  \frac{n\pi x_{0}}{L}\right)  \right\}  \times\\
&  \times\left\{  Q_{2}\left(  \omega\right)  \left(  1+\frac{1}{s_{n}%
}\right)  +Q_{1}\left(  \omega\right)  \left(  1-\frac{1}{s_{n}}\right)
\right\}
\end{align}
where%
\begin{equation}
Q_{1,2}\left(  \omega\right)  =\frac{\eta_{1,2}+i\omega}{\omega^{2}+\eta
_{1,2}^{2}}%
\end{equation}
are the Fourier transforms of $e^{-\eta_{1,2}\tau},$ with $\eta_{1,2}$ given
by (\ref{00}). We see the results for the correlation functions will be given
in terms of Fourier transforms of the arbitrary field variations, similar to
the case with diffusion theory \cite{Clayt}, \cite{Petu}.

This will now be applied to the problem of an axion mediated interaction with
the walls, \cite{Petu} and to the problem of relaxation caused by a magnetic
dipole impurity on the walls of the measuring cell, \cite{Heil}.

\bigskip

\section{Derivation of the false electric dipole moment systematic error in a
Linear Field with the Generalized Conditional Density.}

In the case of a constant gradient across the bound region, the proposed
conditional density should give the same phase shift as \cite{Swank}. Using
the equation (\ref{01})
\[
\left\langle h_{x}\left(  t-\tau\right)  v_{x}(t)\right\rangle =\frac
{\partial}{\partial\tau}\left\langle h_{x}(t)x(t+\tau)\right\rangle
\]

\bigskip Where the correlation of the field strength and position in terms of
the conditional density for a linear gradient,%

\begin{equation}
\left\langle h_{y}(t)y(t+\tau)\right\rangle =\int_{-\frac{L}{2}}^{\frac{L}{2}%
}dx_{0}G_{x}x_{0}~p(x_{0,}t)\int_{-\frac{L}{2}}^{\frac{L}{2}}x~p(x,t~|\;x_{0}%
,t+\tau)~dx
\end{equation}

\bigskip Where $G_{x}$ is the strength of the linear gradient.
\begin{equation}
\left\langle h_{x}(t)x(t+\tau)\right\rangle =\frac{4L^{2}G_{x}}{\pi^{4}%
}e^{-\frac{\tau}{2\tau_{c}}}\sum_{n=odd}^{-\infty\rightarrow\infty}\frac
{1}{n^{4}}\left\{  \cosh\left(  \frac{s_{n}}{2\tau_{c}}\tau\right)  +\frac
{1}{s_{n}}\sinh\left(  \frac{s_{n}}{2\tau_{c}}\tau\right)  \right\}
\end{equation}

\bigskip%
\begin{equation}
\left\langle h_{x}\left(  t-\tau\right)  v_{x}(t)\right\rangle =\frac
{8L^{2}G_{x}}{\pi^{4}}\sum_{n=odd}^{-\infty\rightarrow\infty}\frac{\omega
_{n}^{2}\tau_{c}}{n^{4}s_{n}}e^{-\frac{\tau}{2\tau_{c}}}\sinh\left(
\frac{s_{n}\tau}{2\tau_{c}}\right)
\end{equation}

\bigskip We now write it in a form which only sums $n$ from $1~$to $\infty
.$(we multiply by 2.)$~$%
\[
h_{xv}(\tau)=\left\langle h_{x}\left(  t-\tau\right)  v_{x}(t)\right\rangle
=\frac{16v^{2}\tau_{c}G_{x}}{\pi^{2}}\sum_{n=odd}^{1\rightarrow\infty}\frac
{1}{n^{2}s_{n}}e^{-\frac{\tau}{2\tau_{c}}}\sinh\left(  \frac{s_{n}\tau}%
{2\tau_{c}}\right)
\]

\bigskip To find the phase shift we take the cosine transform according to the
prescription in (\cite{LamGo}) and (\cite{Barab}).%

\begin{align*}
H_{xv}(\omega)  &  =\int_{0}^{\infty}\cos(\omega_{o}\tau)h_{xv}\left(
\tau\right)  d\tau\\
H_{xv}\left(  \omega\right)   &  =\frac{8v^{2}G_{x}}{\pi^{2}}\sum
_{n=odd}^{1\rightarrow\infty}\frac{1}{n^{2}}\frac{\left(  \omega^{2}%
-\omega_{n}^{2}\right)  }{\left(  \omega^{2}-\omega_{n}^{2}\right)  ^{2}%
+\frac{\omega^{2}}{\tau_{c}^{2}}}%
\end{align*}

\bigskip We notice that this field strength correlation spectrum gives the
expected zero frequency value.
\[
H_{xv}\left(  \omega=0\right)  =-G_{x}L_{x}^{2}\frac{8}{\pi^{4}}\sum
_{n=odd}^{1\rightarrow\infty}\frac{1}{n^{4}}%
\]
We arrive at the expected phase shift found in (\cite{LamGo}).\qquad
\qquad\qquad\qquad\qquad\qquad\qquad\qquad\qquad\qquad\qquad\qquad\qquad%
\[
\delta\omega(\omega_{0})=-\frac{\gamma^{2}E}{2c}H_{xv}(\omega_{0})
\]

\bigskip

\section{Appendix, considerations of wall reflections}

We set $t_{0}=0.~$This can be thought of as the time for the RF pulse, where
the conditional density is a delta function and the initial probability
density is flat with value $\frac{1}{L}$. Now we consider a reflection from a
single wall%

\begin{equation}
p(x,t~|~x_{0})~=\delta\left(  x0+vt-\frac{L}{2}-(\frac{L}{2}-x)\right)
=\delta\left(  x0+vt-L+x)\right)
\end{equation}

\bigskip The two terms should be summed for a complete description from $t=0$.
Now say we are confined from -L/2 .. L/2. So we make a second wall reflection
giving us an impulse at
\begin{equation}
p(x,t~|~x_{0})=\delta(x0+vt-2L-x)
\end{equation}

and another wall%

\begin{equation}
p(x,t~|~x_{0})=\delta\left(  x_{0}+vt-3L+x)\right)
\end{equation}

\bigskip and another... \qquad%
\begin{equation}
p(x,t~|~x_{0})=\delta(x_{0}+vt-4L-x)
\end{equation}

More Generally

\ for $v>0$ \ and $x>x_{0}|_{t_{0}=0}$%

\begin{equation}
p(x,t~|~x_{0},t_{0}=0)=\sum_{n=0}^{\infty}\delta(x_{0}+vt-2nL-x)+\sum
_{n=0}^{\infty}\delta(x_{0}+vt-(2n+1)L+x)
\end{equation}

and for $v>0|_{t=t_{0}}$ and $x<x_{0}|_{t_{0}=0}$%

\begin{equation}
p(x,t~|~x_{0},0)=\sum_{n=0}^{\infty}\delta(x_{0}+vt-2nL-x)+\sum_{n=0}^{\infty
}\delta(x_{0}+vt-(2n+1)L+x)
\end{equation}

and for $v<0|_{t=t0}$ and $x>x_{0}|_{t_{0}=0}$%
\begin{equation}
p(x,t~|~x_{0},0)=\sum_{n=0}^{\infty}\delta(x_{0}+vt+2nL+x)+\sum_{n=0}^{\infty
}\delta(x_{0}+vt+(2n+1)L-x)
\end{equation}

\bigskip and for $v<0|_{t=t0}$ and $x<x_{0}|_{t_{0}=0}$%

\begin{equation}
p(x,t~|~x_{0},0)=\sum_{n=0}^{\infty}\delta(x_{0}+vt+2nL+x)+\sum_{n=0}^{\infty
}\delta(x_{0}+vt+(2n+1)L-x).
\end{equation}

\bigskip\ Now we make the assumption that at every $x_{0}$ there should be
equal probability of a $+v$ and $-v.\ $\ Furthermore we can combine the
$x<x_{0}|_{t_{0}=0}~$\ and $x>x_{0}|_{t_{0}=0}~$initial conditions. \ 

With that we have defined a conditional density for particles in one dimension
with no gas scattering and purely specular wall reflections. We use the
expansion of the delta function as:%

\begin{align}
\delta(x-x_{0})  &  =\frac{1}{L}\sum_{n=even}\cos\left(  \frac{n\pi x}%
{L}\right)  \cos\left(  \frac{n\pi x_{0}}{L}\right) \\
&  +\frac{1}{L}\sum_{n=odd}\sin\left(  \frac{n\pi x}{L}\right)  \sin\left(
\frac{n\pi x_{0}}{L}\right)  .\nonumber
\end{align}

Where $n$ goes from $-\infty~~to~\infty$ \ with even/odd respectively for each sum.
\end{document}